\documentclass[12pt]{article}
\usepackage{amsmath,amssymb}
\usepackage{latexsym}

\newcommand{\comm}[2]{\left[#1,#2\right]}

\def\ie{i.e.,\ }
\def\eg{e.g.,\ }

\begin{document}
\title{Contemplations on Dirac's equation in quaternionic coordinates}
\author{Dirk Schuricht and Martin Greiter\\[1cm]
Institut f\"ur Theorie der Kondensierten Materie,\\
  Universit\"at Karlsruhe, Postfach 6980, D-76128 Karlsruhe}
\date{August 17, 2004}
\maketitle
\pagestyle{plain}
\begin{abstract}
   A formulation of Dirac's equation using complex-quaternionic
   coordinates appears to yield an enormous gain in formal elegance,
   as there is no longer any need to invoke Dirac matrices. This
   formulation, however, entails several peculiarities, which we
   investigate and attempt to interpret.\\[-3pt] 
\end{abstract}

\section{Introduction}

One of the breakthroughs in the development of quantum mechanics was
the discovery of a linear, relativistic wave equation for fermions by
Dirac in 1928, which is known as Dirac's equation~\cite{dirac,bjo}.
This equation identified the electron spin as an intrinsic quantum
number and predicted the existence of antiparticles. The corresponding
Lagrangian is at the basis of quantum electrodynamics, and hence has
played a major role on our path towards an understanding of elementary
particles and their interactions. Dirac derived the equation, in a
sense, by ``taking the square root'' of the Klein--Gordon equation,
which is quadratic in time and space derivatives and the minimal
version of a relativistically covariant wave equation. In order to
overcome the technical obstacles in writing a linear equation, Dirac
proposed that the wave equation should be considered as a matrix
equation for complex spinor-valued wave functions rather than complex
scalar functions. The equation contains a set of $4\times 4$ matrices,
the so-called Dirac matrices, for which different representations
satisfying the so-called Dirac algebra are possible.

To the student who learns about these matrices for the first time,
however, they may look somewhat arbitrary. One can easily derive the
algebra and choose a specific representation, but nonetheless, on
purely aesthetic grounds these matrices remain peculiar. A rather
provocative statement by Einstein comes to mind: ``When judging a
physical theory, I ask myself whether I would have made the Universe
in that way had I been God''. In the context of Dirac's equation,
there is little sense in even questioning the correctness of the
theory, but one may still ask whether there might not be a way to
formulate it without availing oneself of these peculiar matrices.

The structure of the matrices in Dirac's equation is intimately linked
to relativistic covariance. Special relativity, however, can not only
be formulated using the standard Lorentz four-vector notation, but
also using a complex-quaternionic (CQ) parametrisation of space-time,
as we have emphasised in a pedagogical review
recently~\cite{GreiterSchurichtejp03}. This parametrisation yields a
significant gain in formal elegance, as Lorentz transformations are no
longer implemented by multiplication with $4\times 4$ matrices, but by
multiplication with CQ numbers. Maxwell's four equations reduce to a
single CQ equation.

The question we wish to address in this article is whether one can
formulate Dirac's equation (and quantum electrodynamics) in CQ
coordinates, and to which extent such a formulation is equivalent to,
or can be brought into a form equivalent to the standard theory. We
will find that relativistic covariance implies immediately that the
entities in the now two-component Dirac spinors are no longer complex
numbers, but CQ numbers, and hence somewhat reminiscent of
quaternionic quantum mechanics~\cite{adler}. Among the peculiarities
intrinsic to this formulation is an apparent doubling of
solutions~\cite{edmonds,rotelli,deleo1}, which we attribute to the
existence of an additional, global $SU(2)$ gauge symmetry, as well as
the possibility to write solutions which carry spin one-half but do
not possess a direction for the quantisation of this spin, which we
eliminate from the spectrum of physical states through a suitable
condition.

Finally, we conclude that while it is indeed possible to gain an
intriguing amount of elegance by formulating Dirac's equation in CQ
coordinates, we have either failed to recognise a deeper physical
principle at work or the price one has to pay for this gain in
elegance is significant.

\section{Special relativity}

We begin with a review of quaternions and the formulation of
special relativity within the framework of CQ 
numbers~\cite{GreiterSchurichtejp03}. We introduce a
complex algebra with generators $1,@ \in \mathbb{C}$, such that
\begin{equation}
@^2=-1
\end{equation}
as well as a quaternionic algebra with generators $1, i, j, k
\in\mathbb{H}$, such that
\begin{equation}
\begin{split}
&i^2=j^2=k^2=-1,\\
&ij=-ji=k,\quad jk=-kj=i,\quad ki=-ik=j,
\end{split}
\end{equation}
which mutually commute:
\begin{equation}
\comm{i}{@}=\comm{j}{@}=\comm{k}{@}=0.
\end{equation}
An arbitrary CQ number can be written as
\begin{displaymath}
q \equiv @t + ix +jy +kz
\end{displaymath}
with $t,x,y,z \in \mathbb{C}$. We further introduce a complex
conjugate operation $^*$ which takes
\begin{displaymath}
@ \rightarrow @^*=-@
\end{displaymath}
but leaves $i$, $j$, and $k$ unchanged, as well as a quaternionic
conjugate operation $^{-}$, which leaves @ unchanged but takes
\begin{displaymath}
i\rightarrow\bar i=-i,\quad j\rightarrow\bar j=-j,\quad \hbox{and}\quad
k\rightarrow \bar k=-k.
\end{displaymath}
Note that if
$ o_1,o_2 \in\mathbb{C}\otimes\mathbb{H}\,$ are two CQ numbers, the
order of the product $o_1o_2$ is reversed under quaternionic
conjugation only:
\begin{equation}
(o_1o_2)^*=o_1^*o_2^*\qquad\hbox{but}\qquad\overline{o_1o_2}=\bar o_2\bar o_1.
\end{equation}
We further define a trace on the complex quaternions via 
\begin{equation}
\mathrm{tr}: \mathbb{C}\otimes\mathbb{H} \rightarrow \mathbb{C}, \ q
\mapsto \frac{1}{2}\left( q+\bar q \right), 
\label{eq:tracedefinition} 
\end{equation} 
which is $\mathbb{C}$-linear and satisfies
\begin{equation}
\mathrm{tr}(pq)=\mathrm{tr}(qp)
\label{eq:traceproperty}
\end{equation}
for $p,q \in\mathbb{C}\otimes\mathbb{H}$.

We label space-time (and other Lorentz contravariant quantities
usually denoted by four vectors) by a purely imaginary CQ number
\begin{equation}
q\equiv @t+ix+jy+kz
\label{eq:im}
\end{equation}
with $t,x,y,z \in \mathbb{R}$, identify this subspace with Minkowski
space, and denote it by $\mathbb{M}$. Complex conjugation $^*$ and
quaternionic conjugation $^{-}$ then correspond to time reversal (T)
and parity (P) transformations, respectively.

The corresponding covariant quantity is given by its quaternionic
conjugate or parity reversed CQ number,
\begin{equation}
\bar q= @t-ix-jy-kz,
\end{equation}
yielding the proper time interval
\begin{equation}
-\bar q q = - q \bar q = t^2-x^2-y^2-z^2.
\label{eq:propertime}
\end{equation}
Defining a scalar product on $\mathbb{M}$
\begin{eqnarray}
\langle p,q\rangle&\equiv &\frac{1}{4}
\left(\overline{(p-q)}(p-q)-\overline{(p+q)}(p+q)\right) \nonumber\\
&=&-\frac{1}{2}(\bar pq+\bar qp) \nonumber\\
&=&-\frac{1}{2}(p\bar q+q\bar p),
\label{eq:scp}
\end{eqnarray}
we find $-\bar q q=\langle q,q\rangle$. With $p=@E+ip_x+jp_y+kp_z$ and
$q=@t+ix+jy+kz$, we have
\begin{equation}
\langle p,q\rangle = Et-p_xx-p_yy-p_zz.
\end{equation}
Note that $@,i,j,k$ form an orthonormal basis of
$\mathbb{M}$.

Let $n=in_x+jn_y+kn_z$ with $n \bar n = n_x^2+n_y^2+n_z^2=1$ and
$n_x,n_y,n_z\in\mathbb{R}$ be a quaternionic imaginary unit
vector. Then a Lorentz transformation is simply given by
\begin{equation}
q\rightarrow q'=\omega q\bar\omega^*,
\label{eq:lt}
\end{equation}
where
\begin{equation}
\omega= e^{\frac{1}{2}n\theta} =
\cos\frac{\theta}{2}+n\sin\frac{\theta}{2}
\label{eq:rot}
\end{equation}
for a rotation by an angle $\theta$ around $n$, or
\begin{equation}
\omega= e^{\frac{1}{2} n @ \Lambda} =
\cosh\frac{\Lambda}{2}+@n\sinh\frac{\Lambda}{2}
\label{eq:boo}
\end{equation}
for a boost by a Lorentz angle $\Lambda$ in direction $n$.

Clearly the covariant CQ number $\bar q$ transforms as
\begin{equation}
\bar q\rightarrow \bar q'=\omega^* \bar q\bar\omega.
\end{equation}
With $\omega \bar \omega = \bar\omega\omega=1$, the Lorentz invariance
of the scalar product $\langle p,q\rangle$ is evident.

\section{Dirac equation} \label{seq:deq}

We now apply the complex-quaternionic formulation of special
relativity to relativistic wave equations. We begin with the simplest
case, \ie a free scalar field. Let $\Phi(q)$ be a function of one
purely imaginary CQ number $q\in\mathbb{M}$, which transforms as a scalar
under Lorentz transformations:
\begin{equation}
\Phi (q)\rightarrow \Phi' (q') = \Phi (q).
\end{equation}
The contravariant differentiation operator is defined by
\begin{equation} 
D \equiv @ \frac{\partial}{\partial t}
-i\frac{\partial}{\partial x} - j\frac{\partial}{\partial y} -
k\frac{\partial}{\partial z} = @\partial_t -i\partial_x -j\partial_y
-k\partial_z
\label{eq:D} 
\end{equation}
and transforms according to
\begin{equation}
D\rightarrow D'=\omega D\bar\omega^*.
\end{equation}
Note that $-\bar D q=4$, which is equivalent to $\partial_\mu x^\mu=4$
in standard notation. With the Lorentz invariant operator $-\bar D D$
the free Klein--Gordon equation reads
\begin{equation}
(-\bar D D + m^2) \Phi(q) = 0,
\label{eq:kge}
\end{equation}
where the real scalar $m$ represents the particle mass. The generally
known solutions of (\ref{eq:kge}) may be written as
\begin{equation}
\Phi (q) = e^{@ \langle p,q\rangle},
\label{eq:kgsolution}
\end{equation}
where the energy-momentum $p=@E+ip_x+jp_y+kp_z$ satisfies
\begin{equation}
-\bar pp = E^2 - p_x^2-p_y^2-p_z^2 = m^2.
\end{equation}

\noindent
In principle, one could also write down solutions of the form
\begin{displaymath}
\Phi (q) = e^{n \langle p,q\rangle},
\end{displaymath}
where $n$ is an arbitrary quaternionic imaginary unit vector. These
solutions, however, do not correspond to any physically new states and, in
contrast to (\ref{eq:kgsolution}), are not ordinary complex
functions. We hence give them no further consideration.

Within this framework, Dirac's equation for the two-component field
$(\psi_1,\psi_2)^\mathrm{T}$ is given by
\begin{equation}
\left( \begin{array}{cc} -m&D \\ \bar D &-m \end{array}\right)
\left( \begin{array}{c} \psi_1 \\ \psi_2 \end{array} \right) =0.
\label{eq:deq}
\end{equation}
This equation is covariant if $m$ is a real scalar and the component
fields transform like
\begin{equation}
\psi_1'=\omega\psi_1 \quad \mathrm{and} \quad \psi_2'=\omega^*\psi_2.
\label{eq:spinortr} 
\end{equation}
This shows that $\psi_{1,2}$ have to be CQ-valued in
general. Therefore, Lorentz trans\-formations are represented by
diagonal matrices (\ref{eq:spinortr}), which is clearly a
simplification compared to the usual spinor representations used in
Dirac's theory~\cite{bjo}.

Iteration of (\ref{eq:deq}) yields the Klein--Gordon equation
(\ref{eq:kge}) for every component $\psi_{1,2}$ and therefore the
relativistic dispersion
\begin{displaymath}
E^2=\vec{p}\;^2+m^2,
\end{displaymath}
where $\vec{p}$ is the spatial momentum. From (\ref{eq:spinortr})
for rotations one finds that, as $\omega (2\pi) =-1$, particles
obeying (\ref{eq:deq}) have spin one-half.

Solving Dirac's equation in the rest frame ($\vec{p}=0$) we obtain the
solutions
\begin{equation}
\Psi =\left(\begin{array}{c} \psi\\ \pm\psi\end{array}\right) e^{\mp @mt},
\quad\psi\in \mathbb{C}\otimes\mathbb{H},
\label{eq:rfs}
\end{equation}
where the upper sign holds for particles ($E>0$) and the lower sign
for antiparticles ($E<0$). If one regards
$\mathbb{C}\otimes\mathbb{H}$ as a four-dimensional complex vector
space, (\ref{eq:rfs}) represents eight linearly independent solutions
of (\ref{eq:deq}), \ie twice the number as for the ordinary Dirac
equation~\cite{bjo}. This doubling of solutions has led
Edmonds~\cite{edmonds} and Gough~\cite{gough} to investigate the
possibility of the existence of hidden quantum numbers and associated
observables. De Leo~\cite{deleo1} proposed that one may reduce the
number of solutions by transforming (\ref{eq:deq}) into a
one-component equation which then requires left and right
multiplications by quaternions. We show in Section~6 below that there
exists an additional symmetry of (\ref{eq:deq}), which reduces the
number of linearly independent solutions from eight to four through
gauge invariance.

Solutions with finite velocity are obtained by writing the exponent in
(\ref{eq:rfs}) as a Lorentz scalar. The Ansatz
\begin{equation} 
\Psi_{\pm}=\left(\begin{array}{c}\varphi_{\pm}(p)\\\xi_{\pm}(p)\end{array}
\right)e^{\mp @\langle p,q\rangle} 
\end{equation}
yields the linear system of equations 
\begin{equation}  
\left( \begin{array}{cc}
m&\pm @p\\ \pm@\bar p&m \end{array}\right) \left(
\begin{array}{c}\varphi_{\pm} (p)\\ \xi_{\pm} (p) \end{array}\right)=0.
\end{equation} 
This implies the spinor condition 
\begin{equation}
\xi_{\pm}(p)=\mp\frac{1}{m}@\;\bar p\;\varphi_{\pm}(p)
\label{eq:spinorcondition}
\end{equation} 
and the energy-momentum relation
\begin{equation}
p\bar p +m^2=-E^2 +\vec{p}\;^2 +m^2 =0.
\end{equation}
Note that (\ref{eq:spinorcondition}) is invariant under Lorentz
transformations and is satisfied by (\ref{eq:rfs}).

\section{Spin and spin directions}

In the previous section we have seen that the solutions of
(\ref{eq:deq}) describe particles with spin one-half. For further
investigations note that, as the spin operator is the generator of
rotations in the rest frame, we have to restrict ourselves to
particles in their rest frame, which are represented by states of the
form (\ref{eq:rfs}). Let us first identify the components of the spin
operator. Consider a rotation around the $x$-axis by an angle
$\theta$. With $\omega=\omega^*$ we obtain
\begin{equation}
\omega \psi_{1,2}=e^{\frac{1}{2}i\theta}\psi_{1,2}=
e^{-@\frac{@}{2}i\theta}\psi_{1,2}.
\label{eq:rotseries}
\end{equation}
The components of the spin operator are now easily read off to be 
\begin{equation}
S_x\equiv\frac{@}{2}i,\quad S_y\equiv\frac{@}{2}j,\quad\mathrm{and}\quad
S_z\equiv\frac{@}{2}k.
\label{eq:spinoperators}
\end{equation}
With $\comm{i}{j}=2k$ and cyclic permutations one readily verifies
that (\ref{eq:spinoperators}) satisfy the angular momentum algebra
\begin{equation}
\comm{S_x}{S_y}=@S_z,\quad\comm{S_y}{S_z}=@S_x,\quad\comm{S_z}{S_x}=@S_y.
\label{eq:spinalgebra}
\end{equation}
For the square of the spin operator we obtain $\vec S^2=\frac{3}{4}$,
as expected. The spin operator $\vec{S}$ is hence implemented through
a simple left multiplication with the quaternionic basis elements. Our
construction differs from that of De Leo~\cite{deleo1} and
Gough~\cite{gough2}, who proposed operators consisting of simultaneous
left and right multiplication with $i,j,k$.

Let us now choose the $z$-axis as spin quantisation direction. The
solutions of (\ref{eq:deq}) which are simultaneously eigenstates of
$S_z$ with eigenvalues $m_z=\pm \frac{1}{2}$ are given by
\begin{equation}
\begin{array}{rrl}
m_z=&\!\frac{1}{2}:&\ (1+@k)\Psi_0,\ (i+@j)\Psi_0 \\
\rule{0pt}{15pt}m_z=&\!\!-\frac{1}{2}:&\ (@i+j)\Psi_0,\ (-@-k)\Psi_0,
\end{array}
\label{eq:spinev}
\end{equation}
where  
\begin{displaymath} 
\Psi_0=\left(\begin{array}{c} 1\\1 \end{array}\right)e^{-@mt}
\end{displaymath}
for particles. Antiparticle solutions are constructed similarly.
With the usual definition for the raising and lowering operators
$S^{\pm}=S_1 \pm @S_2$ one easily verifies that the two subspaces 
\begin{equation}
\mathrm{span}_{\mathbb{C}}\{(1+@k)\Psi_0,(@i+j)\Psi_0\} \quad
\mathrm{and} \quad
\mathrm{span}_{\mathbb{C}}\{(i+@j)\Psi_0,(-@-k)\Psi_0\}
\label{eq:spinsubspaces}
\end{equation}
are closed under the spin algebra (\ref{eq:spinalgebra}). Note that we
can convert the spin eigenvectors (\ref{eq:spinev}) for the same
eigenvalue into each other by right multiplication with quaternionic
basis elements, \eg
\begin{equation}
(1+@k)\Psi_0 i = (i+@j)\Psi_0 \quad\mathrm{and}\quad (@i+j)\Psi_0i=(-@-k)\Psi_0.
\end{equation}
The right multiplication with the other basis elements $j,k$ does not
yield independent states. Hence we can connect the two spin
eigenspaces (\ref{eq:spinsubspaces}) by right multiplication with
$i,j,k$. We will return to this issue in Section~6.

Another peculiarity of the CQ formulation of Dirac's equation is the
following. The states
\begin{equation} 
\Psi_0,\; i\Psi_0,\; j\Psi_0,\; k\Psi_0
\label{eq:dsbasis}
\end{equation}
form a basis of the subspace of particle solutions. They are somewhat
peculiar, however, as they mix the two spin eigenspaces
(\ref{eq:spinsubspaces}), \eg we have
\begin{equation}
\Psi_0=\frac{1}{2}(1+@k)\Psi_0+\frac{@}{2}(-@-k)\Psi_0.
\label{eq:psi_0decomposition}
\end{equation}
Note that the eigenvectors (\ref{eq:spinev}) of $S_z$ satisfy the
condition $\Psi^\mathrm{T}\bar\Psi=0$ (where $^\mathrm{T}$ denotes
transposition) whereas the states (\ref{eq:dsbasis}) do not. In
general, the following two statements are equivalent:
\begin{equation}
\Psi^\mathrm{T} \bar\Psi =0\quad\Longleftrightarrow\quad 
\left(\vec{e}\cdot\vec{S}\right)\Psi=\pm\frac{1}{2}\Psi,
\label{eq:equivalence}
\end{equation}
where $\vec e$ is a unit vector in $\mathbb{R}^3$. The dot $\cdot$
denotes the standard scalar product. 
The right hand side of (\ref{eq:equivalence}) states that the spin of
the particle points in the direction $\vec{e}$. The states
(\ref{eq:dsbasis}) are solutions to Dirac's equation and possess spin
one-half, but in contrast to the usual situation there is no direction
$\vec{e}$ of the spin. Rewriting (\ref{eq:equivalence}) for the states
(\ref{eq:rfs}) we obtain the condition
\begin{equation}
\psi\bar\psi=0 \quad\Longleftrightarrow\quad
\left(\vec{e}\cdot\vec{S}\right)\psi=\pm\frac{1}{2}\psi.
\label{eq:psicondition}
\end{equation}
With $\psi=\psi_0+i\psi_1+j\psi_2+k\psi_3$, the equation on the left
can be rewritten
\begin{equation}
\psi_0^2+\psi_1^2+\psi_2^2+\psi_3^2=0.
\label{eq:quadric}
\end{equation}
Since $\mathbb{C}\otimes\mathbb{H}$ is isomorphic to the space of
complex $2\times 2$ matrices~\cite{waerden}, (\ref{eq:quadric}) is
equivalent to the statement that $\psi$ is not invertible in
$\mathbb{C}\otimes\mathbb{H}$. The condition (\ref{eq:psicondition})
holds for antiparticles as well.

To prove (\ref{eq:equivalence}) we start with the eqivalent condition
(\ref{eq:psicondition}). First note that $\vec e\cdot\vec
S=\frac{@}{2}n$ with $n\in\mathbb{H}\backslash\mathbb{R}$. Then
multiplying (\ref{eq:psicondition}) by $\bar\psi$ from the right and
assuming $\psi\bar\psi\neq0$, we obtain the contradiction
\begin{displaymath}
\pm\frac{1}{2}=\frac{@}{2}n\not\in\mathbb{R}.
\end{displaymath}
Hence, in order to fulfil (\ref{eq:psicondition}) we need $\psi\bar\psi=0$.

The other direction is proven by rewriting
$\psi\in\mathbb{C}\otimes\mathbb{H}$ as
\begin{equation}
\psi=a+@b+cm+@dm'
\label{eq:standardform}
\end{equation}
with $m,m'$ quaternionic imaginary unit vectors and
$a,b,c,d\in\mathbb{R}$. Then, by setting $b=0$, we have the condition 
\begin{displaymath}
\psi\bar\psi=(a+cm)(a-cm)-(a+cm)@dm'+@dm(a-cm)-d^2=0,
\end{displaymath}
which yields
\begin{equation}
a^2+c^2-d^2=0,\quad (mm'+m'm)cd=0.
\end{equation}
For $\psi\neq 0$ the equation on the left implies $d\neq 0$.  In the
first case we have $c=0$ and can choose $n=m'$ in
(\ref{eq:psicondition}), in the second case we have $mm'+m'm=0$ and we
can choose $n=\pm\frac{1}{d}(a+cm)m'$, which completes the proof.

The above implies that states obeying (\ref{eq:equivalence}) may be
interpreted as ordinary spin one-half particles with usual spin
properties. There are, however, two subspaces (\ref{eq:spinsubspaces})
closed under the spin algebra (\ref{eq:spinalgebra}) rather than one,
which can be connected by right multiplication with the quaternionic
basis elements. In addition to these familiar states there are also
states with $\Psi^\mathrm{T}\bar\Psi\neq 0$ which are not
interpretable as ordinary particles. We will return to these
peculiarities in Section~6.

\section{Quantum electrodynamics}

So far we considered only free particles. We now turn to Dirac
fermions coupled to an external electromagnetic potential, which we
describe with the Lagrangian formalism. For free Dirac fermions, the
Lorentz invariant Lagrangian density is given by
\begin{eqnarray}
\mathcal{L}_0 &=&
\mathrm{tr}\left(\left( \begin{array}{cc} \bar\psi_2^*&\!\!\!,\;\bar\psi_1^*
\end{array} \right)
\left( \begin{array}{cc} -m&D\\ \bar D &-m\end{array}\right)
\left( \begin{array}{c} \psi_1 \\ \psi_2 \end{array} \right)\right)\nonumber \\
&=&\rule{0pt}{20pt} \mathrm{tr}\left(\bar\psi_1^* \bar D\psi_1 +
\bar\psi_2^* D\psi_2 -m\bar\psi_1^*\psi_2-m\bar\psi_2^*\psi_1 \right),
\label{eq:deq-lag}
\end{eqnarray}
where the trace is as defined in (\ref{eq:tracedefinition}). The free
action is given as usual by the integral
\begin{equation}
\mathcal{S}_0 = \int_{\mathbb{M}} \mathcal{L}_0 \ d^4\!q,
\label{eq:deq-action}
\end{equation}
where $d^4\!q$ denotes the volume element in $\mathbb{M} \cong \mathbb{R}^4$.
With $D^*=-\bar D$, $\bar D^* =-D$ we find 
\begin{equation}
\mathcal{L}_0^*=\mathrm{tr}\left(-\bar\psi_1 D\psi_1^*
-\bar\psi_2 \bar D\psi_2^*-m\bar\psi_1\psi_2^*-m\bar\psi_2\psi_1^*\right),
\end{equation}
which with integration by parts yields the reality condition
$\mathcal{S}_0^* =\mathcal{S}_0$. The trace in (\ref{eq:deq-lag})
implies $\bar{\mathcal{S}}_0=\mathcal{S}_0$ such that
$\mathcal{S}_0\in\mathbb{R}$.

To couple the Dirac fields to the electromagnetic potential we require
invariance under the local gauge transformation 
\begin{equation}
\Psi \rightarrow\Psi'=\Psi e^{-@e\alpha (q)}
\label{eq:u1gauge}
\end{equation}
with $e$ a real constant and $\alpha : \mathbb{M} \rightarrow
\mathbb{R}$ a scalar function. This implies the minimal coupling
procedure $D\rightarrow D+@eA$ in (\ref{eq:deq}) and the
transformation rules for the vector field $A\equiv
@\phi+iA_x+jA_y+kA_z$:
\begin{eqnarray}
A&\rightarrow &\omega A\bar\omega^* \hspace{0.83 cm} \mathrm{under\enspace
Lorentz\enspace transformations},\\
A&\rightarrow &A + D\alpha \quad \mathrm{under}\enspace U(1)\enspace
\mathrm{gauge\enspace transformations}.
\end{eqnarray}
The interaction is hence given by the Lagrangian density 
\begin{equation}
\mathcal{L}_\mathrm{int}=\mathrm{tr}\left(@e\bar\psi_1^*\bar A\psi_1
+@e\bar\psi_2^* A\psi_2\right),
\label{eq:int-lag}
\end{equation}
which is real as $A^*=-\bar A$ and $\bar A^*=-A$. We identify $e$ with
the electric charge. The electromagnetic field strength for $A$ is
given by~\cite{GreiterSchurichtejp03}
\begin{equation}
F\equiv \frac{1}{2}\left(\bar DA-\overline{\bar DA}\right),
\label{eq:F}
\end{equation}
and the Lagrangian density for the free electromagnetic field is
\begin{equation}
\mathcal{L}_{A}=\frac{1}{4}\left(F^2+(F^*)^2\right),
\label{eq:A-lag}
\end{equation}
which is real since $\bar F=-F$. The Lagrangian density of quantum
electrodynamics is hence
\begin{equation}
\mathcal{L}_\mathrm{QED}=\mathcal{L}_0+\mathcal{L}_\mathrm{int}+
\mathcal{L}_A.
\end{equation}
From the non-relativistic limit of Dirac's equation, we can further
read off the $g$-factor of the electron to equal $g=2$.

\section{Quaternionic gauge invariance and the connection to the conventional 
Dirac equation}\label{seq:quatgauge}

We have seen in Section~3 that a doubling of solutions occurs in the
CQ version of Dirac's equation. This is also evident from the
existence of two closed spin eigenspaces (\ref{eq:spinsubspaces})
rather than one. We have seen that these two eigenspaces are connected
by right multiplication with $i,j,k$.

We generalize these transformations to
\begin{equation} 
\Psi \rightarrow \Psi'=\Psi e^{-n\beta},
\label{eq:quaternionicgauge}
\end{equation}
where $n=in_x+jn_y+kn_z$ is a quaternionic imaginary unit vector and
$\beta \in \mathbb{R}$. Note that $e^{-n\beta}$ can be represented in
the basis $1,i,j,k$ of $\mathbb{H}$ (see also (\ref{eq:rot})). The
Lagrangian density (\ref{eq:deq-lag}) is, by use of
(\ref{eq:traceproperty}), invariant under this transformation. This
is, however, not a physical symmetry of the system but rather an
invariance of description within the CQ formulation of Dirac's
equation, as we will explain now.

Note first that the unit quaternion $e^{-n\beta}$ represents a point
on the three sphere $\mathcal{S}^3\subset\mathbb{R}^4$. Therefore,
(\ref{eq:quaternionicgauge}) represents a global $SU(2)$ gauge
symmetry. The standard procedure to introduce a gauge symmetry is to
require the system to be invariant under a local gauge
transformation. To meet this requirement one introduces a gauge field
(like $A$ for the $U(1)$ gauge transformations in Section~5) which is
minimally coupled to the matter field $\Psi$. This coupling represents
the interaction of the matter with the gauge field. The extension of
(\ref{eq:quaternionicgauge}) to a local gauge transformation, \ie
$\beta\equiv\beta(q)$, however, does not seem possible within the CQ
formulation of Dirac's equation presented here. To see this, consider
for simplicity $\beta(q)=x$ and introduce a gauge field $B$ by
replacing the differentiation operator $D$ by $D+B$. Then we have to
require
\begin{equation}
((D+B)\Psi)'\stackrel{!}{=}(D+B)'\Psi'=D\Psi'+B'\Psi'
\end{equation}
under the (local) transformation (\ref{eq:quaternionicgauge}). For
$\beta(q)=x$ this yields
\begin{equation}
(D\Psi)e^{-nx}+B\Psi e^{-nx}=(D\Psi)e^{-nx}+i\Psi n e^{-nx}+B'\Psi e^{-nx},
\label{eq:qgaugeproblem}
\end{equation}
where we have used the definition of the differentiation operator $D$
(see (\ref{eq:D})). As $n\in\mathbb{H}$ and there is no general rule
to express $\Psi n$ in terms of $n\Psi$, it is impossible to satisfy
(\ref{eq:qgaugeproblem}) with $B$ a CQ-valued function or
matrix~\cite{reference}.

It is hence not possible to elevate the global ``gauge'' symmetry
(\ref{eq:quaternionicgauge}) to a true and local gauge symmetry. The
symmetry is, however, a gauge symmetry, as it corresponds to an
invariance of our description (as gauge symmetries generally do)
rather than an invariance of the physical system. It is significant
for the interpretation of the CQ formulation of Dirac's equation, as
it connects the two spin eigenspaces (\ref{eq:spinsubspaces}) but
leaves the spin eigenvalues invariant. Identifying states connected by
quaternionic gauge transformations (\ref{eq:quaternionicgauge}) with
each other, the number of particle solutions of the free Dirac
equation (\ref{eq:deq}) reduces from two to one for each spin
eigenvalue. Note that (\ref{eq:quaternionicgauge}) is trivially
extended to quantum electrodynamics with $A \rightarrow A$.

Recalling (\ref{eq:psi_0decomposition}), we see that states with
$\Psi^\mathrm{T}\bar\Psi\neq 0$ correspond to superpositions of states
formulated in different quaternionic gauges. The crucial point is that
no gauge transformation exists which transforms a state with
$\Psi^\mathrm{T}\bar\Psi\neq 0$ into a state which belongs to only one
of the subspaces (\ref{eq:spinsubspaces}), \eg there is no solution
for $n$ and $\beta$ satisfying
\begin{equation}
\Psi_0e^{-n\beta}=a(1+@k)\Psi_0+b(@i+j)\Psi_0,\quad a,b\in\mathbb{C}.
\end{equation} 
By contrast, spin eigenstates can be transformed by
(\ref{eq:quaternionicgauge}) to states which belong to any one of the
spin eigenspaces (\ref{eq:spinsubspaces}). For example,
\begin{equation}
\frac{1}{\sqrt{2}}(1+@k)\Psi_0+\frac{1}{\sqrt{2}}(i+@j)\Psi_0=
(1+@k)\Psi_0e^{i\frac{\pi}{4}}.
\end{equation}
This does, however, not provide us with a definite answer to the
question whether states with $\Psi^\mathrm{T}\bar\Psi\neq 0$ should be
interpreted physically or should be excluded from the theory by
requiring $\Psi^\mathrm{T}\bar\Psi=0$ for physical states
explicitly. The contemplations summarized above suggest the latter.

To verify that there are no further gauge symmetries, note that the
Lagrangian density (\ref{eq:deq-lag}) is invariant under right
multiplication of a CQ number $\Sigma$ only if
$\Sigma\bar\Sigma^*=1$. Then the statement
\begin{equation}
  \Sigma\bar\Sigma^*=1\Rightarrow\Sigma=cq,
\label{eq:normal-cq}
\end{equation}
where $c\in\mathbb{C}$, $q\in\mathbb{H}$, and $|c||q|=1$, shows that
there are no gauge symmetries besides the $U(1)$ symmetry as described
in Section~5 and the quaternionic gauge transformations
(\ref{eq:quaternionicgauge}). The proof of (\ref{eq:normal-cq}) is
straightforward, but somewhat to technical for this article.

We conclude this section by relating the CQ version of Dirac's
equation (\ref{eq:deq}) to the standard formulation with complex
matrices~\cite{bjo}. For this purpose, we expand the CQ spinors
$\psi_{1,2}$ in up- and down-spin components, where we can restrict
ourselves due to (\ref{eq:quaternionicgauge}) (at least for states
satisfying $\Psi^\mathrm{T}\bar\Psi=0$) to the first of the two
subspaces (\ref{eq:spinsubspaces}), \ie we write
\begin{equation}  
\psi_1=(1+@k)c_1+(@i+j)c_2,\quad\psi_2=(1+@k)c_3+(@i+j)c_4
\end{equation} 
with $c_i : \mathbb{M}\rightarrow\mathbb{C}$. Inserting this in 
(\ref{eq:deq}) yields four independent equations which combine to 
\begin{equation}  
(@\gamma^{\mu}\partial_{\mu}-m)\mathcal{C}=0 \quad \mathrm{with} \quad
\mathcal{C}=(c_1,c_2,c_3,c_4)^\mathrm{T},  
\end{equation}
where the gamma matrices are given in the chiral representation:
\begin{equation}  
\gamma^0=\left(\begin{array}{cc}\mathbf{0}&\mathbf{1}\\
    \mathbf{1}&\mathbf{0}\end{array}\right),\quad
\gamma^i=\left(\begin{array}{cc}\mathbf{0}&\mathbf{\sigma^\mathit{i}}\\
    \mathbf{-\sigma^\mathit{i}}&\mathbf{0}\end{array} \right),
\label{eq:diracmatrices}
\end{equation}
with $\mathbf{0}$ and $\mathbf{1}$ are the $2\times 2$ zero and unit
matrix and $\mathbf{\sigma^\mathit{i}}$ the Pauli matrices.  The CQ
formulation with $\Psi^\mathrm{T}\bar\Psi=0$ presented here is hence
equivalent to the standard complex formulation of Dirac's equation.

\section{Conclusion}

In this article, we have extended the complex-quaternionic formulation
of special relativity~\cite{GreiterSchurichtejp03} to Dirac's
equation, obtaining an equation for a two-component CQ field. In this
formulation there is no need to invoke Dirac matrices, and Lorentz
transformations are given by diagonal matrices rather than complicated
spinor representations. The price to pay for these simplifications is,
however, outrageous. First, we encountered a doubling of solutions, a
problem which we overcame by attributing them to different gauge
choices of a global $SU(2)$ gauge symmetry. Second, we encountered
particle solutions with spin one-half but without a spin quantisation
direction. According to conventional quantum mechanics these solutions
appear to be unphysical, but this did not emerge from the formalism we
developed above. This led us to derive a condition to ban them from
the Hilbert space, which we imposed {\it a posteriori}. Furthermore,
there are mathematical problems inherent to the formulation due to the
fact that $\mathbb{C}\otimes\mathbb{H}$ is not a division
algebra~\cite{adler,dixon}.

In summary, we found no indication that the CQ formulation of Dirac's
equation would lead to a deepening of our understanding of
relativistic fer\-mi\-ons. In our opinion, the problems we encountered
with this formulation eradicate the gain in formal elegance we were
hoping to obtain.

\section*{Acknowledgment}
One of us (DS) was supported by the German Research Foundation (DFG)
through GK 284.

\end{document}